\documentstyle[twocolumn,prl,aps]{revtex}
\begin{document}
\draft
\title{Comment on ``Theory of Spinodal Decomposition''}
\author{A. D. Rutenberg\cite{newaddress}}
\address{Department of Physics and Astronomy,
The University of Manchester, M13 9PL, UK}

\maketitle
\date{\today}

The basis of Goryachev's analysis \cite{Goryachev94} of conserved
scalar phase-ordering dynamics, to apply only the global constraint
$\int \psi d {\bf x} = \text{const.}$, is incorrect.
For physical conserved systems, which evolve by mass transport,
the stronger local conservation law embodied by the continuity equation
\begin{equation}
\label{EQ:CONT}
	\partial \psi /\partial t + \nabla \cdot {\bf j} =0,
\end{equation}
is the appropriate one to use \cite{Gunton83}. Even as an approximation, the
global constraint is inadequate \cite{Tamayo89}.

The standard evolution equation for systems with conserved dynamics is
\begin{equation}
\label{EQ:DYN}
     \partial \psi / \partial t =  \nabla^2 \delta F/ \delta \psi,
\end{equation}
 where
$F[\psi] = \int d {\bf x} \left[ (\nabla \psi)^2 + V_0 (\psi^2-1)^2 \right]$
is the effective free energy. These dynamics satisfy the local conservation
law (\ref{EQ:CONT}), and are motivated {\em phenomenologically} by a current
${\bf j} = - \nabla \delta F/ \delta \psi$.  At very early times after a
quench from a disordered state, gradients will be large and higher order
gradient terms will be needed.  Other disagreements with (\ref{EQ:DYN}) can
stem, for example, from hydrodynamic, thermal, and stress relaxation effects.
These indicate important extensions needed to (\ref{EQ:DYN}) and $F[\psi]$,
however the local conservation (\ref{EQ:CONT}) will still apply in all of these
cases.

A special initial condition emphasizes the differences in the microscopic
evolution of local vs. global conservation, where we
only require that the dissipative dynamics be invariant under
$\psi \rightarrow -\psi$ and that $F[\psi]$
is minimized by $\psi = \pm 1$ everywhere. Consider two half
spaces, antisymmetric about a static flat domain wall,
one of which has $\psi=1$ everywhere except for a small sphere where
$\psi=-1$, the other of which has $\psi=+1$ and $-1$ respectively.
For spheres far from the domain wall, under local
conserved dynamics the total magnetization of each half space will be constant
as the spheres evolve. However with only global conservation,
always satisfied by the symmetry of the problem,
the dynamics are identical to non-conserved dynamics and the magnetization
of each half-space will evolve in time and will eventually saturate.
This is clearly inconsistent with a local conservation law.

The differences between the global constraint and a local conservation
law is also made clear by a class of dynamics introduced by Onuki
\cite{Onuki85} that includes both cases. In Fourier space we have
\begin{equation}
\label{EQ:GEN}
   \partial \psi_{\bf k}/\partial t = - | {\bf k}|^\sigma \delta F/\delta
    \psi_{-{\bf k}},
\end{equation}
where $\sigma =2$ is the locally conserved  dynamics of
(\ref{EQ:DYN}), $\sigma \rightarrow 0^+$ imposes the global constraint
discussed by Goryachev, and $\sigma=0$ is non-conserved dynamics.
The differences between local and global conservation
laws can be clearly seen in the late time behavior after a quench, which must
be governed by the same non-linear dynamics as the early-time behavior.
As discussed in a unified treatment \cite{Bray94} of (\ref{EQ:GEN}), and
in agreement with previous results \cite{Gunton83}, the growth laws are
$L(t) \sim t^{1/3}$ for (locally) conserved scalar quenches, and
$L(t) \sim t^{1/2}$ for non-conserved and globally constrained quenches,
where $t$ is the time since the quench.  $L(t)$ also describes the radius of
the spheres in the previous paragraph, evolving by (\ref{EQ:GEN}), where $t$
is the time to annihilation.

We can also consider long-range interactions within the
effective free-energy $F[\psi]$. These are
relevant both for attractive \cite{Bray94} and for repulsive \cite{Sagui95},
or competing, interactions.  The free-energy should enter into the
dynamics the same way, independently of any long-range interactions.
This leads to similar differences between local conservation and a global
constraint.

Any approximate treatment must start from dynamics that are
phenomenologically consistent with microscopic dynamical processes
and from effective free energies that are consistent with equilibrium
properties. It is incorrect for Goryachev to apply only a global constraint to
represent physical systems with local conservation laws.

I thank A. J. Bray for discussions and the Isaac Newton Institute for
hospitality.

\end{document}